\documentclass[letterpaper, conference]{IEEEtran}
\IEEEoverridecommandlockouts

\ifCLASSOPTIONcompsoc
  \usepackage[nocompress]{cite}
\else
  \usepackage{cite}
\fi
\ifCLASSINFOpdf
\else
\fi

\usepackage{amssymb}
\usepackage{multirow}
\usepackage{multicol}
\usepackage{url}
\usepackage{array}
\usepackage{verbatim}
\usepackage{float}
\usepackage{units}
\usepackage{textcomp}
\usepackage{amsmath}
\usepackage[utf8]{inputenc}
\usepackage{graphicx}
\usepackage{caption}
\usepackage{subcaption}
\usepackage{setspace}
\usepackage{placeins}
\usepackage{rotating}
\usepackage{mdwmath}
\usepackage{varwidth}
\usepackage{algpseudocode}

\def\BibTeX{{\rm B\kern-.05em{\sc i\kern-.025em b}\kern-.08em
    T\kern-.1667em\lower.7ex\hbox{E}\kern-.125emX}}

\newcommand{\MYfooter}{\smash{
\hfil\parbox[t][\height][t]{\textwidth}{\centering
\thepage}\hfil\hbox{}}}

\makeatletter

\def\ps@headings{%
\def\@oddhead{\parbox[t][\height][t]{\textwidth}{\centering
UAV-Based RGB and Multispectral Indices for Palm Tree Precision Agriculture\\
}\hfil\hbox{}}%

\def\@evenhead{\parbox[t][\height][t]{\textwidth}{\centering
\\
}\hfil\hbox{}}%

\def\@oddfoot{\MYfooter}%
\def\@evenfoot{\MYfooter}}

\def\ps@IEEEtitlepagestyle{%
\def\@oddhead{\parbox[t][\height][t]{\textwidth}{\centering
}\hfil\hbox{}}%
\def\@evenhead{\scriptsize\thepage \hfil \leftmark\mbox{}}%
\def\@oddfoot{  \textcopyright \hfil 
\leftmark\mbox{}}%
\def\@evenfoot{\MYfooter}}

\makeatother
\begin{document}
%

\title{Evaluation of UAV-Based RGB and Multispectral Vegetation Indices for Precision Agriculture in Palm Tree Cultivation}

\author
{\IEEEauthorblockN{Alavikunhu Panthakkan$^{1, a}$, S M Anzar$^{2}$, K. Sherin$^{3}$, Saeed Al Mansoori $^{4}$ and Hussain Al-Ahmad$^{5}$}
\IEEEauthorblockA{\textit{$^{1 \& 5}$College of Engineering and IT, University of Dubai, U.A.E.}\\
\textit{$^{2}$Department of Electronics and Communication, TKM College of Engineering, Kollam, India-691 005}\\
\textit{$^{2}$ Department of Electronics and Communication, MES College of Engineering, Kuttippuram, India-679 582}\\
\textit{$^{4}$ Applications Development and Analysis Section, Mohammed Bin Rashid Space Centre (MBRSC), UAE}\\
\textit{Corresponding Author: $^{a}$apanthakkan@ud.ac.ae}
} \\}


%


\maketitle
\IEEEpubid{\begin{minipage}{\textwidth}\ \\ \\[12pt] 
\normalsize 979-8-3503-4524-7/23/\$31.00
\copyright2023 IEEE. 
\end{minipage}} 
\begin{abstract}
Precision farming relies on accurate vegetation monitoring to enhance crop productivity and promote sustainable agricultural practices. This study presents a comprehensive evaluation of UAV-based imaging for vegetation health assessment in a palm tree cultivation region in Dubai. By comparing multispectral and RGB image data, we demonstrate that RGB-based vegetation indices offer performance comparable to more expensive multispectral indices, providing a cost-effective alternative for large-scale agricultural monitoring. Using UAVs equipped with multispectral sensors, indices such as NDVI and SAVI were computed to categorize vegetation into healthy, moderate, and stressed conditions. Simultaneously, RGB-based indices like VARI and MGRVI delivered similar results in vegetation classification and stress detection. Our findings highlight the practical benefits of integrating RGB imagery into precision farming, reducing operational costs while maintaining accuracy in plant health monitoring. This research underscores the potential of UAV-based RGB imaging as a powerful tool for precision agriculture, enabling broader adoption of data-driven decision-making in crop management. By leveraging the strengths of both multispectral and RGB imaging, this work advances the state of UAV applications in agriculture, paving the way for more efficient and scalable farming solutions.
\end{abstract}

\IEEEpeerreviewmaketitle

\section{Introduction}
Precision farming has become a pivotal approach in modern agriculture, aimed at boosting crop productivity, ensuring sustainable practices, and optimizing the use of resources~\cite{getahun2024application}. A key component of precision farming is the accurate and timely monitoring of vegetation, which is essential for assessing plant health, detecting stress, and facilitating data-driven decisions in crop management\cite{paul2022viable}. Traditional vegetation monitoring techniques often rely on labor-intensive fieldwork, making them both time-consuming and costly. However, with the advent of Unmanned Aerial Vehicles (UAVs) equipped with advanced imaging sensors, the landscape of precision farming has shifted. UAV-based remote sensing offers a more efficient, scalable, and accurate method for gathering vital crop health data over large areas.

Multispectral sensors, widely used in UAVs, have been extensively researched for their ability to provide detailed information about plant health~\cite{barbedo2019review}. Multispectral imaging is commonly employed to calculate vegetation indices like the Normalized Difference Vegetation Index (NDVI) and the Soil Adjusted Vegetation Index (SAVI), which are useful in classifying vegetation into healthy, moderate, and stressed categories~\cite{qi1994modified}. Despite their effectiveness, the high cost of multispectral sensors often limits their widespread application, especially in large-scale farming operations or regions with financial constraints.
\begin{figure*}[ht]
    \centering
    \includegraphics[scale=0.5]{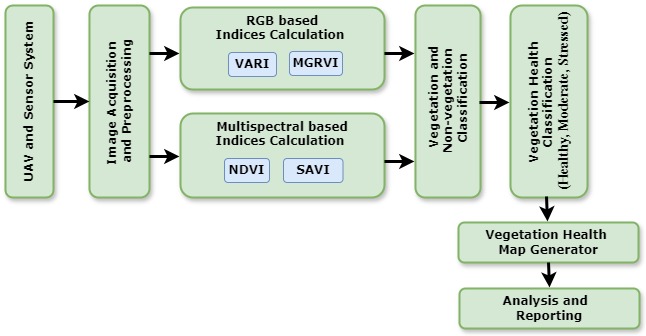}
    \caption{Flowchart of the methodology for UAV-based vegetation health assessment using RGB and multispectral indices.}
    \label{fig:block_diagram}
\end{figure*}
This study is motivated by the need to explore more cost-effective alternatives to multispectral imaging while maintaining the accuracy required for vegetation monitoring. RGB cameras, which are more affordable and widely available, have shown promising potential for providing comparable vegetation indices. Indices like the Visible Atmospherically Resistant Index (VARI) and the Modified Green Red Vegetation Index (MGRVI) have demonstrated the ability to capture critical information about plant health using simple RGB imagery. Given the significant cost difference between RGB and multispectral sensors, understanding whether RGB-based vegetation indices can deliver comparable accuracy is essential for promoting the broader adoption of UAV technology in agriculture, especially in resource-limited environments.

The primary objective of this study is to evaluate the performance of RGB-based vegetation indices compared to traditional multispectral indices within the context of precision farming. Specifically, we focus on UAV-based monitoring of palm tree cultivation in Dubai, where both multispectral indices (NDVI, SAVI) and RGB-based indices (VARI, MGRVI) are computed to assess vegetation health. By comparing the performance of these indices, this research aims to demonstrate that RGB-based monitoring provides a cost-effective alternative to multispectral imaging without compromising accuracy. The findings of this study have the potential to improve the scalability and affordability of precision farming technologies, allowing for more widespread implementation of sustainable agricultural practices.

\section{Literature Review}

Unmanned Aerial Vehicles (UAVs) have become integral tools in precision agriculture, enabling detailed, efficient, and cost-effective monitoring of vegetation health. UAVs equipped with both multispectral and RGB sensors provide high-resolution imagery, allowing for the calculation of various vegetation indices to assess plant health, stress levels, and growth dynamics.

Multispectral sensors, commonly used in UAV applications, have been widely studied for computing indices like the \textit{Normalized Difference Vegetation Index (NDVI)} and the \textit{Soil-Adjusted Vegetation Index (SAVI)}. NDVI is one of the most commonly used indices for assessing vegetation health by exploiting the difference between the red and near-infrared (NIR) bands. NDVI values typically range from -1 to 1, with higher values indicating healthier vegetation. However, NDVI can be limited in areas with significant soil exposure, which led to the development of SAVI, a soil brightness correction factor that improves accuracy in arid and semi-arid regions \cite{huete1988soil,dubbini2015evaluating}.

Recent studies have explored the potential of using RGB sensors as a cost-effective alternative to multispectral sensors. \textit{RGB-based vegetation indices}, such as the \textit{Visible Atmospherically Resistant Index (VARI)} and the \textit{Modified Green Red Vegetation Index (MGRVI)}, have been effectively employed in UAV-based monitoring to provide insights into vegetation health. Although RGB sensors do not capture the NIR band, indices like VARI can offer reliable information on vegetation health using visible light data \cite{bendig2015combining,gitelson2022novel}. This makes RGB sensors an attractive option for farmers in resource-constrained environments where multispectral sensors may be too costly.

Comparative analyses between RGB and multispectral indices show that, while multispectral indices provide more detailed insights due to the inclusion of NIR data, RGB indices still perform well in monitoring general vegetation health. For instance, VARI has been found to be robust in detecting greenness and stress in vegetation, especially under different atmospheric conditions, making it suitable for large-scale agricultural monitoring \cite{dubbini2015evaluating}. Additionally, studies on MGRVI have shown its ability to differentiate between healthy and stressed vegetation based on chlorophyll content in the green and red bands \cite{bendig2015combining,isprs2022evaluation}.

Recent research highlights the growing importance of UAVs in precision agriculture, as they offer rapid, flexible, and high-resolution data collection. UAVs equipped with RGB and multispectral sensors have been applied successfully in various crop systems, including rice, vineyards, and palm tree cultivation, demonstrating their potential for wide-scale use \cite{zhang2022comparison}. As UAV technology becomes more accessible, the integration of both RGB and multispectral imagery offers promising solutions for cost-effective and scalable precision farming practices.

While RGB-based indices provide a promising alternative to multispectral indices, particularly in resource-limited environments, more research is needed to refine the accuracy and sensitivity of RGB indices for large-scale agricultural applications. Future studies could explore advanced image processing techniques to further enhance the precision of RGB-based monitoring in different crop systems and environmental conditions.

\begin{figure*}[t!]
    \centering
    \begin{subfigure}[b]{0.45\linewidth}
        \centering
        \includegraphics[scale=0.45]{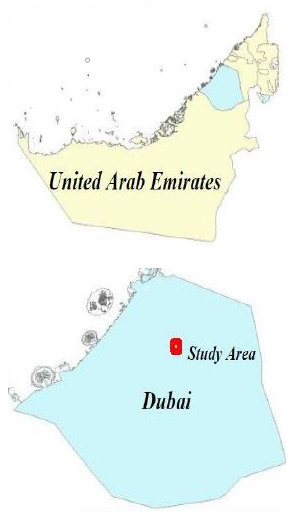}
        \caption{Study Area}
        \label{fig:Study}
    \end{subfigure}
    \hfill
    \begin{subfigure}[b]{0.45\linewidth}
        \centering
        \includegraphics[scale=0.3]{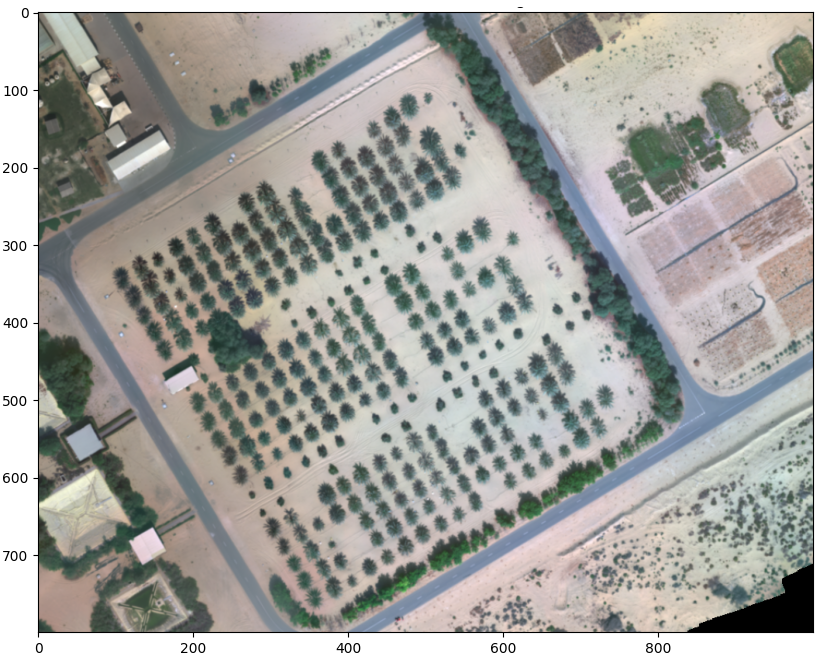}
        \caption{RGB ROI Image}
        \label{fig:SAVI}
    \end{subfigure}
    \caption{Palm Tree Region}
    \label{fig:ndvi_savi_maps}
\end{figure*}

\section{Methodology}

The goal of this study is to evaluate the effectiveness of UAV-based imaging for assessing vegetation health in palm tree cultivation, with a focus on comparing the performance of RGB and multispectral vegetation indices. UAVs equipped with multispectral and RGB sensors were deployed to capture high-resolution images. Vegetation indices including NDVI, SAVI, VARI, and MGRVI were computed from the acquired imagery to assess plant health and classify vegetation into categories such as healthy, moderate, and stressed.

This methodology follows a structured approach, starting with image acquisition, followed by data preprocessing, calculation of vegetation indices, and classification of vegetation health. The study emphasizes a comparative analysis of the cost-effectiveness and accuracy of RGB-based indices versus traditional multispectral indices, particularly in the context of arid regions like Dubai.

The block diagram in Figure~\ref{fig:block_diagram} illustrates the overall workflow of the methodology used for evaluating vegetation health using UAV-based RGB and multispectral vegetation indices in palm tree cultivation. The process starts with a UAV and sensor system, where Unmanned Aerial Vehicles (UAVs) equipped with RGB and multispectral sensors capture high-resolution images of the target area. These images undergo image acquisition and preprocessing, which includes steps to correct noise and enhance image quality, ensuring that the data is ready for accurate analysis.

Once preprocessed, the images are split into two parallel processing streams. The first stream involves RGB-based indices calculation, where indices like VARI (Visible Atmospherically Resistant Index) and MGRVI (Modified Green Red Vegetation Index) are computed. These indices use visible light data to assess vegetation health and offer a cost-effective solution. The second stream is dedicated to multispectral-based indices calculation, where NDVI (Normalized Difference Vegetation Index) and SAVI (Soil-Adjusted Vegetation Index) are computed from multispectral data. These indices utilize the near-infrared spectrum, which is highly sensitive to vegetation health, especially in areas where soil exposure is significant.

After the indices are calculated from both RGB and multispectral data, they undergo vegetation and non-vegetation classification, separating vegetation from non-vegetative elements such as soil, water, or infrastructure. The classified vegetation is then further divided into three categories based on health status: healthy, moderate, and stressed, depending on the index values.

Using the health classification, a vegetation health map is generated, providing a visual representation of the health status across the palm tree cultivation area. Finally, the process concludes with analysis and reporting, where the results are interpreted, and actionable insights are generated, including identifying stressed regions that may require intervention. This methodology leverages both cost-effective RGB data and more sensitive multispectral data to offer a comprehensive and scalable solution for precision agriculture.

\begin{figure*}[t!]
    \centering
    \begin{subfigure}[b]{0.45\linewidth}
        \centering
        \includegraphics[scale=0.45]{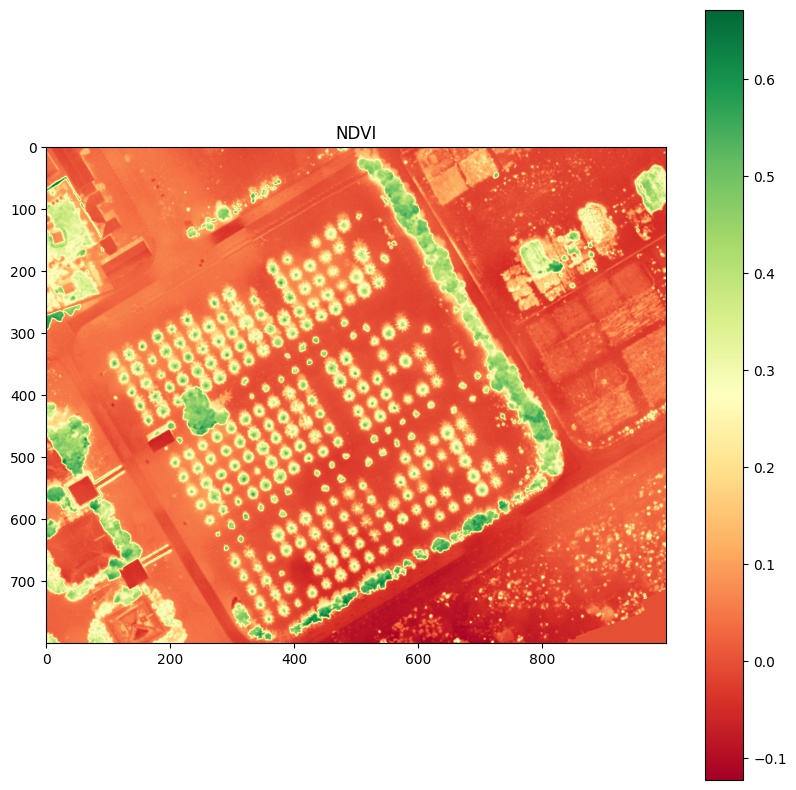}
        \caption{NDVI}
        \label{fig:NDVI}
    \end{subfigure}
    \hfill
    \begin{subfigure}[b]{0.45\linewidth}
        \centering
        \includegraphics[scale=0.45]{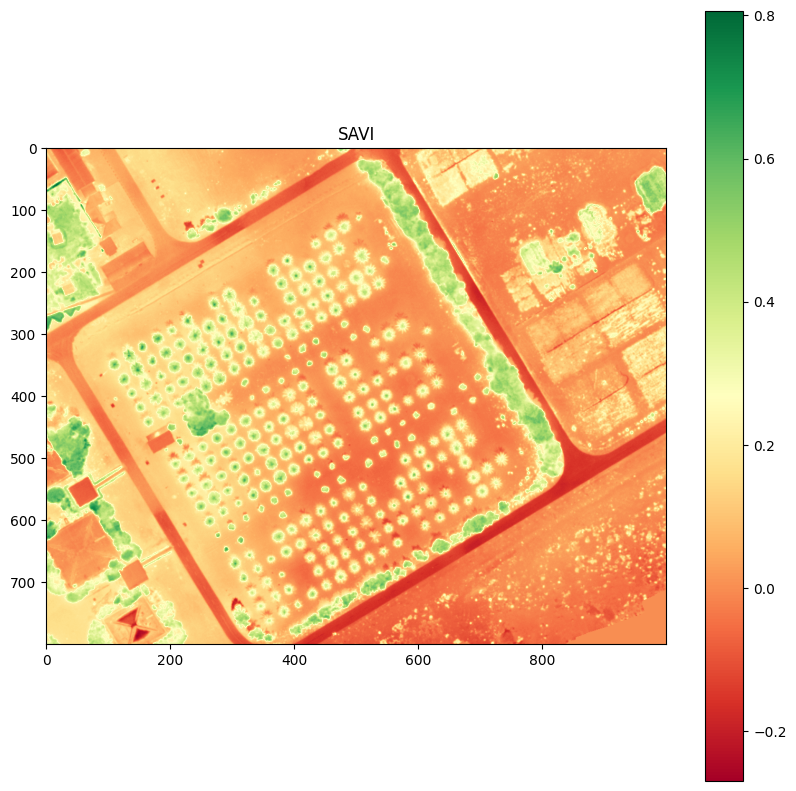}
        \caption{SAVI}
        \label{fig:SAVI}
    \end{subfigure}
    \caption{NDVI and SAVI maps of the selected palm plot}
    \label{fig:ndvi_savi_maps}
\end{figure*}

\begin{figure*}[t!]
    \centering
    \begin{subfigure}[b]{0.45\linewidth}
        \centering
        \includegraphics[scale=0.4]{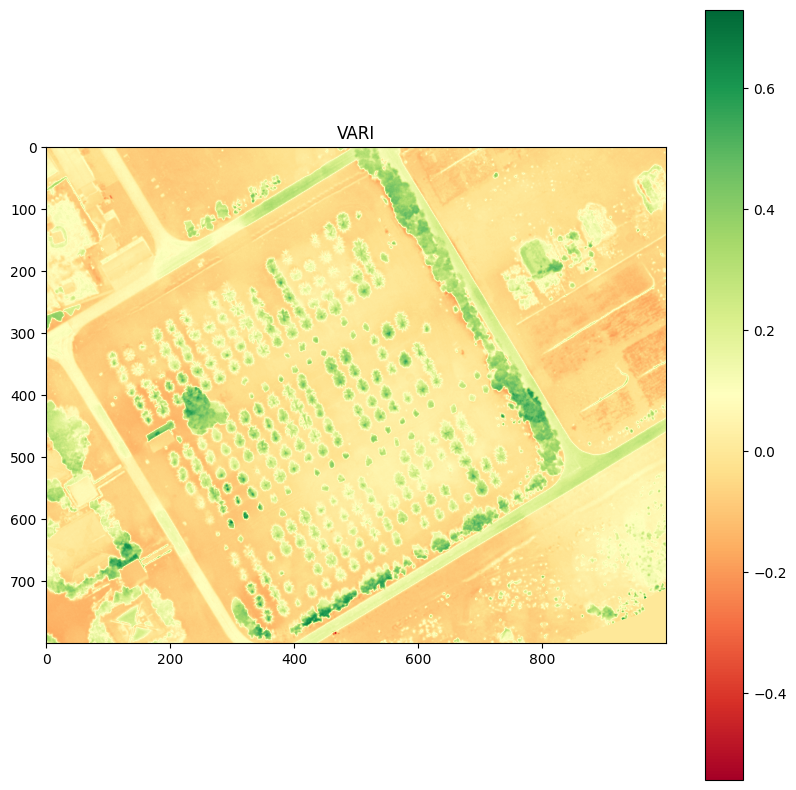}
        \caption{NDVI}
        \label{fig:NDVI}
    \end{subfigure}
    \hfill
    \begin{subfigure}[b]{0.45\linewidth}
        \centering
        \includegraphics[scale=0.4]{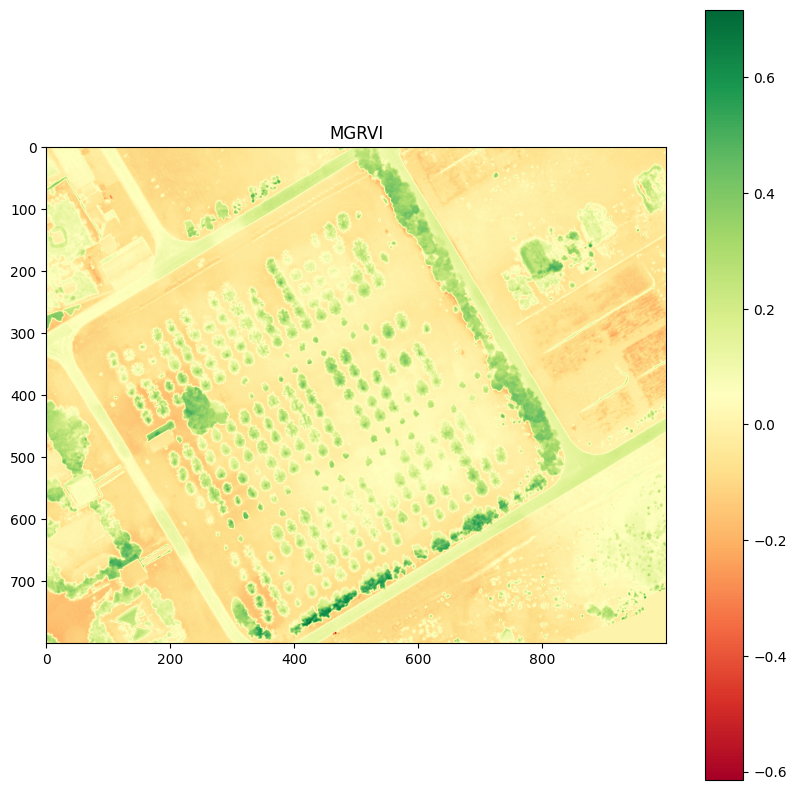}
        \caption{SAVI}
        \label{fig:SAVI}
    \end{subfigure}
    \caption{Visual Representation of Vegetation Index for RGB Image}
    \label{fig:ndvi_savi_maps}
\end{figure*}
\begin{figure*}[t!]
    \centering
    \begin{subfigure}[b]{0.9\linewidth}
        \centering
        \includegraphics[scale=0.5]{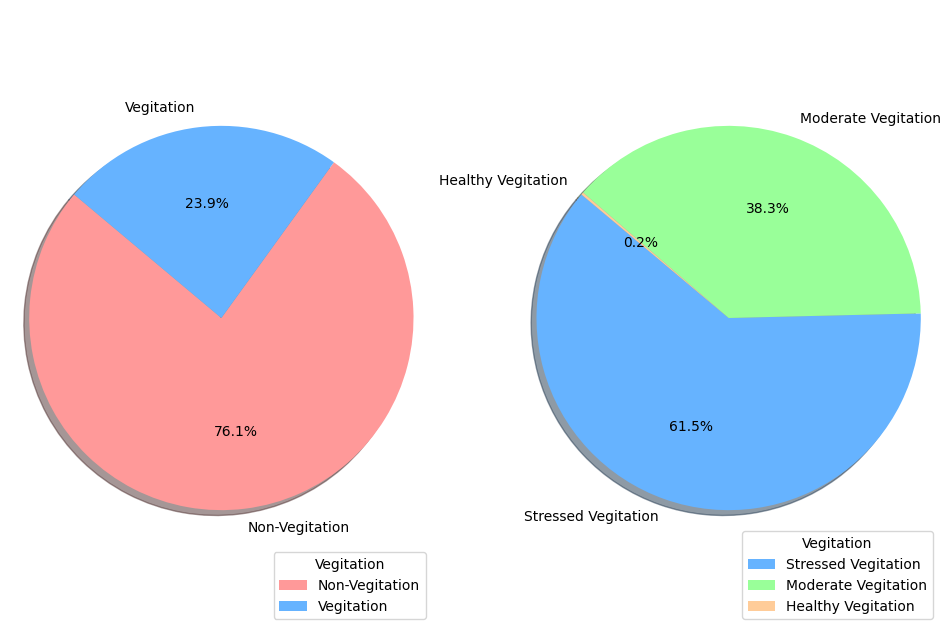}
        \caption{NDVI}
        \label{fig:NDVI}
    \end{subfigure}
    \vfill
    \begin{subfigure}[b]{0.9\linewidth}
        \centering
        \includegraphics[scale=0.5]{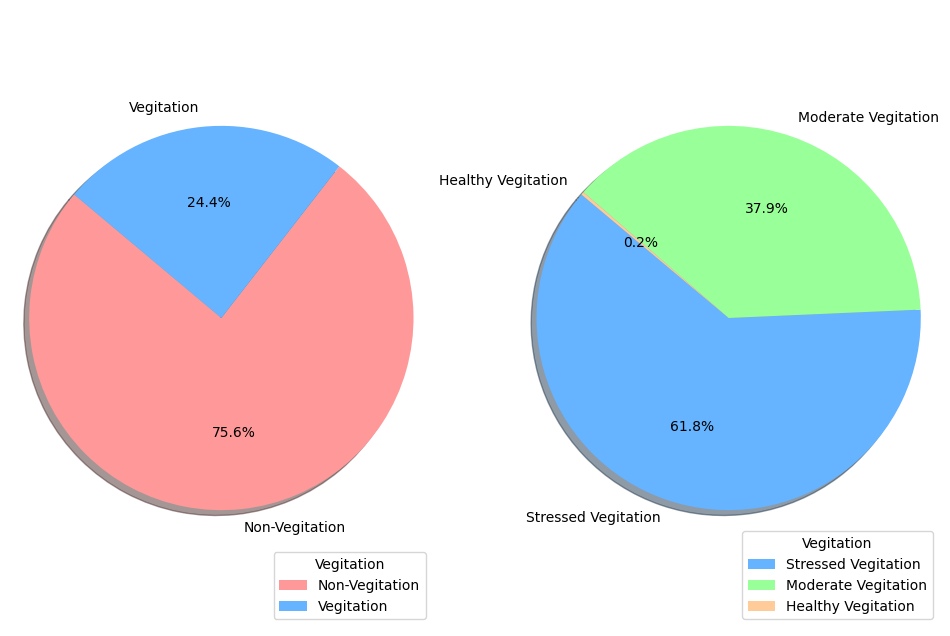}
        \caption{SAVI}
        \label{fig:SAVI}
    \end{subfigure}
    \caption{Pie chart of vegetation indices (NDVI and SAVI) using multispectral images}
    \label{fig:pie_charts_ndvi_savi}
\end{figure*}
\begin{figure*}[t!]
    \centering
    \begin{subfigure}[b]{0.9\linewidth}
        \centering
        \includegraphics[scale=0.5]{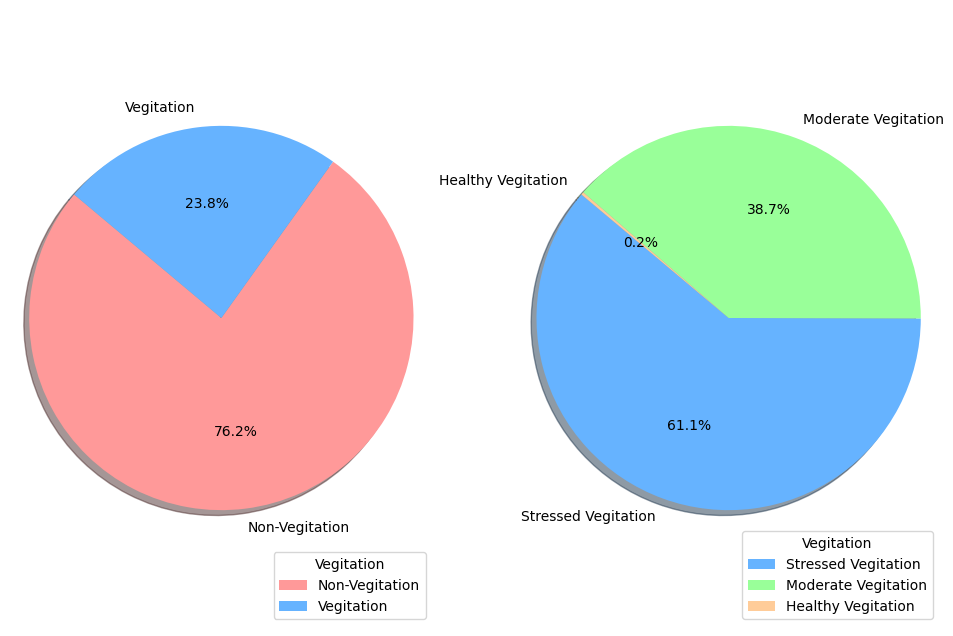}
        \caption{VARI}
        \label{fig:VARI}
    \end{subfigure}
    \vfill
    \begin{subfigure}[b]{0.9\linewidth}
        \centering
        \includegraphics[scale=0.5]{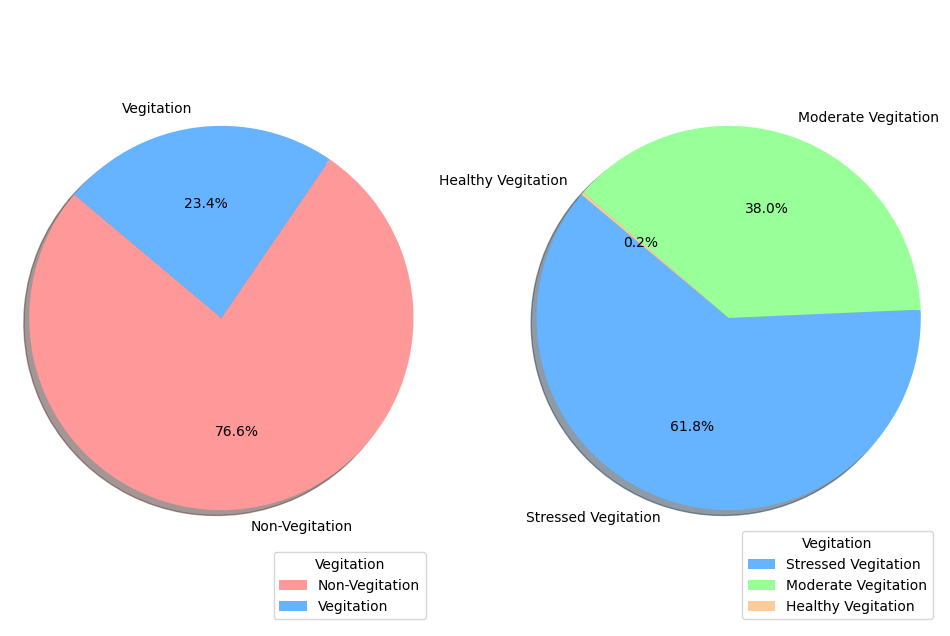}
        \caption{MGRVI}
        \label{fig:MGRVI}
    \end{subfigure}
    \caption{Pie chart of vegetation indices (VARI and MGRVI) of RGB images}
    \label{fig:pie_charts_vari_mgrvi}
\end{figure*}
\begin{table*}[ht]
\centering
\caption{Vegetation indices derived from multispectral imagery, with index ranges and percentages for non-vegetation and vegetation areas}
\begin{tabular}{|c|c|c|c|c|c|c|}
\hline
\multirow{2}{*}{Index} & \multirow{2}{*}{Minimum} & \multirow{2}{*}{Maximum} & \multicolumn{2}{|c|}{Non-Vegetation} & \multicolumn{2}{|c|}{Vegetation} \\ \cline{4-7}
 & & & Range  & Percentage  & Range  & Percentage \\ \hline
NDVI   & -0.1223 & 0.67114 & -1 $\leq$ NDVI $\leq$ 0.1 & 76.15\% & 0.1 \textless NDVI $\leq$ 1 & 23.85\%    \\ \hline
SAVI   & -0.2693 & 0.8061  & -1 $\leq$ SAVI $\leq$ 0.16 & 75.59\% & 0.16 \textless SAVI $\leq$ 1 & 24.41\%    \\ \hline
\end{tabular}
\label{tab:veg_indices}
\end{table*}
\begin{table*}[ht]
\centering
\caption{Vegetation indices derived from multispectral imagery, further subdividing vegetation into stress, moderate, and dense categories with corresponding index ranges and percentages}
\begin{tabular}{|c|c|c|c|c|c|c|}
\hline
\multirow{2}{*}{Index} & \multicolumn{2}{|c|}{Stress Vegetation} & \multicolumn{2}{|c|}{Moderate Vegetation} & \multicolumn{2}{|c|}{Dense Vegetation} \\ \cline{2-7}
 & Range   & Percentage  & Range  & Percentage  & Range & Percentage \\ \hline
NDVI   & 0.1 \textless NDVI $\leq$ 0.3  & 61.49\%  & 0.3 \textless NDVI $\leq$ 0.6  & 38.30\%   & 0.6 \textless NDVI $\leq$ 1 & 0.20\% \\ \hline
SAVI   & 0.16 \textless SAVI $\leq$ 0.33  & 61.81\%  & 0.33 \textless SAVI $\leq$ 0.64 & 37.94\%  & 0.64 \textless SAVI $\leq$ 1 & 0.25\%     \\ \hline
\end{tabular}
\label{tab:veg_indices2}
\end{table*}

\begin{table*}[ht]
\centering
\caption{Vegetation indices derived from RGB imagery, with index ranges and percentages for non-vegetation and vegetation areas}
\begin{tabular}{|c|c|c|c|c|c|c|}
\hline
\multirow{2}{*}{Index} & \multirow{2}{*}{Minimum} & \multirow{2}{*}{Maximum} & \multicolumn{2}{|c|}{Non-Vegetation} & \multicolumn{2}{|c|}{Vegetation} \\ \cline{4-7}
 &  &  & Range                              & Percentage                       & Range                            & Percentage \\ \hline
VARI    & -0.5438     & 0.72972     & -1 $\leq$ VARI $\leq$ 0.08              & 76.24\%                          & 0.08 \textless VARI $\leq$ 1             & 23.76\%    \\ \hline
MGRVI   & -0.614      & 0.7169      & -1 $\leq$ MGRVI $\leq$ 0.08             & 76.58\%                          & 0.08 \textless MGRVI $\leq$ 1            & 23.42\%    \\ \hline
\end{tabular}
\label{tab:veg_indices}
\end{table*}
\begin{table*}[ht]
\centering
\caption{Vegetation indices derived from RGB imagery, further subdividing vegetation into stress, moderate, and dense categories with corresponding index ranges and percentages}
\begin{tabular}{|c|c|c|c|c|c|c|}
\hline
\multirow{2}{*}{Index} & \multicolumn{2}{|c|}{Stress Vegetation} & \multicolumn{2}{|c|}{Moderate Vegetation} & \multicolumn{2}{|c|}{Dense Vegetation} \\ \cline{2-7}
 & Range                              & Percentage    & Range                              & Percentage     & Range                              & Percentage \\ \hline
VARI    & 0.08 \textless VARI $\leq$ 0.22            & 61.08\%                          & 0.22 < VARI $\leq$ 0.59           & 38.70\%                          & 0.59 \textless VARI $\leq$ 1              & 0.23\%    \\ \hline
MGRVI   & 0.08 \textless MGRVI $\leq$ 0.2            & 61.81\%                          & 0.2 < MGRVI $\leq$ 0.59           & 37.97\%                          & 0.59 \textless MGRVI $\leq$ 1             & 0.22\%    \\ \hline
\end{tabular}
\label{tab:veg_indices}
\end{table*}

\subsection{Vegetation Indices}
Vegetation indices are mathematical tools used in remote sensing to assess plant health, growth, and stress by analyzing reflectance in different spectral bands. Common indices like NDVI (Normalized Difference Vegetation Index) and SAVI (Soil-Adjusted Vegetation Index) use near-infrared (NIR) and red bands to monitor vegetation health, with NDVI being the most widely used. SAVI improves upon NDVI by reducing soil brightness effects, making it useful in arid regions. For more affordable RGB-based imaging, indices like VARI (Visible Atmospherically Resistant Index) and MGRVI (Modified Green Red Vegetation Index) provide valuable insights into vegetation health. These indices are crucial in precision farming, allowing for efficient and cost-effective monitoring of crops and guiding resource management. The following sections provide details on the specific vegetation indices used in this study.
\subsubsection{Normalized Difference Vegetation Index (NDVI)}
NDVI is a widely-used metric for assessing vegetation health by leveraging the reflectance differences in the red (R) and near-infrared (NIR) bands. It is calculated as:
\begin{equation}
\text{NDVI} = \frac{NIR - R}{NIR + R}
\end{equation}
Higher NDVI values indicate healthier and denser vegetation, while lower values correspond to stressed or sparse vegetation. NDVI is commonly used in agriculture for monitoring crop health, detecting stress, and optimizing resource management, although it may saturate in areas with high biomass and be influenced by soil reflectance~\cite{huang2021commentary}.
\subsubsection{Soil-Adjusted Vegetation Index (SAVI)}
SAVI was developed to mitigate the influence of soil reflectance in regions with sparse vegetation. It introduces a correction factor \( L \) to reduce soil brightness effects and is calculated as:
\begin{equation}
\text{SAVI} = \frac{(NIR - R) \times (1 + L)}{NIR + R + L}
\end{equation}
Typically, \( L \) is set to 0.5, but it can be adjusted depending on vegetation density. SAVI is particularly useful in arid and semi-arid regions, providing improved accuracy over NDVI in areas with exposed soil~\cite{qi1994modified}.

\subsubsection{Visible Atmospherically Resistant Index (VARI)}
VARI uses only the visible spectrum (RGB) to estimate vegetation fraction, offering robustness to atmospheric variations. It is calculated as:
\begin{equation}
\text{VARI} = \frac{G - R}{G + R - B}
\end{equation}
Where \( G \), \( R \), and \( B \) represent green, red, and blue reflectance values, respectively. VARI is advantageous for applications using standard RGB cameras and is commonly applied in agricultural settings to assess vegetation health in real-time~\cite{anisa2020uav}.
\subsubsection{Modified Green Red Vegetation Index (MGRVI)}
MGRVI is used to differentiate between healthy and stressed vegetation by emphasizing green and red reflectance. It is calculated as:
\begin{equation}
\text{MGRVI} = \frac{G^2 - R^2}{G^2 + R^2}
\end{equation}
Positive values indicate healthy vegetation, while negative values suggest stressed vegetation or bare soil. MGRVI enhances the sensitivity to chlorophyll content and is particularly effective in green-dominated environments.

\subsection{Database and Study Area}
This study was conducted in a palm tree cultivation region in Dubai, UAE, an arid environment where high temperatures and limited water availability pose significant challenges for agriculture. The region’s climatic conditions make precision farming essential for optimizing crop health and productivity, particularly in palm tree cultivation, a critical agricultural activity in the area.

The database for this study consists of UAV-captured imagery from both multispectral and RGB sensors. The UAVs were equipped with sensors capable of acquiring high-resolution images, allowing for the computation of key vegetation indices. Specifically, for multispectral data, the Normalized Difference Vegetation Index (NDVI) and Soil-Adjusted Vegetation Index (SAVI) were calculated, providing insight into vegetation health and soil-adjusted vegetation status. For RGB imagery, the Visible Atmospherically Resistant Index (VARI) and Modified Green Red Vegetation Index (MGRVI) were used to assess vegetation health in a cost-effective manner. These indices enabled the classification of vegetation into healthy, moderate, and stressed conditions, supporting precision farming efforts to optimize resource use and crop management in this challenging environment.

\section{Results and Discussion}
This study evaluated the performance of both RGB and multispectral vegetation indices for assessing vegetation health in palm tree cultivation in Dubai. The indices used include the Normalized Difference Vegetation Index (NDVI) and Soil-Adjusted Vegetation Index (SAVI) for multispectral imagery, and the Visible Atmospherically Resistant Index (VARI) and Modified Green Red Vegetation Index (MGRVI) for RGB imagery. The following results are based on the generated vegetation maps and pie charts, as well as a comparative analysis of the vegetation index values.

\subsection{Vegetation Indices Derived from Multispectral Imagery (NDVI and SAVI)}
Figures 1(a) and 1(b) present the NDVI and SAVI maps for the palm tree plot, with Table I summarizing the percentage distribution of vegetation and non-vegetation areas. NDVI values range from -0.1223 to 0.67114, where 76.15\% of the area is classified as non-vegetation ($NDVI \leq 0.1$) and 23.85\% as vegetation ($NDVI > 0.1$). For SAVI, the range is from -0.2693 to 0.8061, with 75.59\% non-vegetation ($SAVI \leq 0.16$) and 24.41\% vegetation ($SAVI > 0.16$).

Further subdivision of the vegetation into stress, moderate, and dense categories is shown in Table II. For NDVI, 61.49\% of the vegetation was classified as stressed ($0.1 < NDVI \leq 0.3$), 38.30\% as moderate ($0.3 < NDVI \leq 0.6$), and only 0.20\% as dense ($NDVI > 0.6$). Similarly, for SAVI, 61.81\% was stressed ($0.16 < SAVI \leq 0.33$), 37.94\% moderate ($0.33 < SAVI \leq 0.64$), and 0.25\% dense ($SAVI > 0.64$).

These results indicate that the majority of the vegetation in the study area is under stress, with a small portion categorized as healthy or dense vegetation.

\subsection{Vegetation Indices Derived from RGB Imagery (VARI and MGRVI)}
Figures 2(a) and 2(b) show the vegetation maps for the RGB-based indices, VARI and MGRVI. Table III provides data on vegetation and non-vegetation areas. For VARI, the index ranges from -0.5438 to 0.72972, with 76.24\% of the area classified as non-vegetation ($VARI \leq 0.08$) and 23.76\% as vegetation ($VARI > 0.08$). Similarly, for MGRVI, the range is from -0.614 to 0.7169, with 76.58\% non-vegetation ($MGRVI \leq 0.08$) and 23.42\% vegetation ($MGRVI > 0.08$).

Table IV further categorizes vegetation into stress, moderate, and dense categories. For VARI, 61.08\% of the vegetation was stressed ($0.08 < VARI \leq 0.22$), 38.70\% moderate ($0.22 < VARI \leq 0.59$), and 0.23\% dense ($VARI > 0.59$). For MGRVI, 61.81\% was stressed ($0.08 < MGRVI \leq 0.2$), 37.97\% moderate ($0.2 < MGRVI \leq 0.59$), and 0.22\% dense ($MGRVI > 0.59$).

\subsection{Discussion}
The comparative analysis between multispectral (NDVI and SAVI) and RGB-based (VARI and MGRVI) vegetation indices reveals consistent patterns of vegetation stress across both datasets. A large portion of the vegetation was categorized as stressed, with a smaller percentage classified as moderate or dense. This indicates that RGB-based indices provide results comparable to multispectral indices, making them a cost-effective alternative for large-scale agricultural monitoring.

While RGB indices are generally less sensitive than multispectral ones, they still offer valuable insights into vegetation health. The use of RGB-based indices, such as VARI and MGRVI, is particularly advantageous in resource-limited environments where cost-effective monitoring is critical. However, for more precise assessments, especially in areas with dense vegetation, multispectral indices like NDVI and SAVI may offer better accuracy due to their ability to capture near-infrared information.

Overall, the study demonstrates that integrating RGB imagery into precision farming provides a scalable, affordable solution for vegetation monitoring, particularly in challenging environments such as the arid regions of Dubai.

\section{Conclusion}
\label{sec:Conclusion}
\par This study demonstrated that UAV-based RGB vegetation indices, such as VARI and MGRVI, offer a cost-effective alternative to multispectral indices like NDVI and SAVI for monitoring vegetation health in palm tree cultivation. Both RGB and multispectral indices provided comparable results in classifying vegetation into stressed, moderate, and dense categories.

While multispectral indices offer greater sensitivity due to near-infrared data, RGB-based indices proved sufficient for meaningful vegetation assessments, particularly in resource-limited environments. This highlights the potential for affordable RGB-based solutions in precision agriculture, reducing costs without compromising accuracy.

This work encourages the broader adoption of UAV-based imaging for efficient and scalable vegetation monitoring, particularly in challenging environments like the arid regions of Dubai. Future research can explore enhancing the precision of RGB indices and their applicability across different crops and environments.


\begin{thebibliography}{10}
\providecommand{\url}[1]{#1}
\csname url@samestyle\endcsname
\providecommand{\newblock}{\relax}
\providecommand{\bibinfo}[2]{#2}
\providecommand{\BIBentrySTDinterwordspacing}{\spaceskip=0pt\relax}
\providecommand{\BIBentryALTinterwordstretchfactor}{4}
\providecommand{\BIBentryALTinterwordspacing}{\spaceskip=\fontdimen2\font plus
\BIBentryALTinterwordstretchfactor\fontdimen3\font minus \fontdimen4\font\relax}
\providecommand{\BIBforeignlanguage}[2]{{%
\expandafter\ifx\csname l@#1\endcsname\relax
\typeout{** WARNING: IEEEtran.bst: No hyphenation pattern has been}%
\typeout{** loaded for the language `#1'. Using the pattern for}%
\typeout{** the default language instead.}%
\else
\language=\csname l@#1\endcsname
\fi
#2}}
\providecommand{\BIBdecl}{\relax}
\BIBdecl

\bibitem{getahun2024application}
S.~Getahun, H.~Kefale, and Y.~Gelaye, ``Application of precision agriculture technologies for sustainable crop production and environmental sustainability: A systematic review,'' \emph{The Scientific World Journal}, vol. 2024, no.~1, p. 2126734, 2024.

\bibitem{paul2022viable}
K.~Paul, S.~S. Chatterjee, P.~Pai, A.~Varshney, S.~Juikar, V.~Prasad, B.~Bhadra, and S.~Dasgupta, ``Viable smart sensors and their application in data driven agriculture,'' \emph{Computers and Electronics in Agriculture}, vol. 198, p. 107096, 2022.

\bibitem{barbedo2019review}
J.~G.~A. Barbedo, ``A review on the use of unmanned aerial vehicles and imaging sensors for monitoring and assessing plant stresses,'' \emph{Drones}, vol.~3, no.~2, p.~40, 2019.

\bibitem{qi1994modified}
J.~Qi, A.~Chehbouni, A.~R. Huete, Y.~H. Kerr, and S.~Sorooshian, ``A modified soil adjusted vegetation index,'' \emph{Remote sensing of environment}, vol.~48, no.~2, pp. 119--126, 1994.

\bibitem{huete1988soil}
A.~Huete, ``A soil-adjusted vegetation index (savi),'' \emph{Remote Sensing of Environment}, vol.~25, no.~3, pp. 295--309, 1988.

\bibitem{dubbini2015evaluating}
M.~Dubbini, M.~Gattelli, and R.~Scopigno, ``Evaluating multispectral images and vegetation indices for precision farming applications from uav images,'' \emph{Remote Sensing}, vol.~7, no.~4, pp. 4026--4047, 2015.

\bibitem{bendig2015combining}
J.~Bendig, A.~Bolten, S.~Bennertz, J.~Broscheit, S.~Eichfuss, and G.~Bareth, ``Combining uav-based plant height from crop surface models, visible, and near-infrared vegetation indices for biomass monitoring in barley,'' \emph{International Journal of Applied Earth Observation and Geoinformation}, vol.~39, pp. 79--87, 2015.

\bibitem{gitelson2022novel}
A.~Gitelson, D.~Rundquist, and Y.~Kaufman, ``Vegetation indices derived from remote sensing to assess crop growth and yield,'' \emph{Agricultural and Forest Meteorology}, vol. 112, no.~1, pp. 15--20, 2022.

\bibitem{isprs2022evaluation}
``Evaluation of rgb vegetation indices derived from uav images for rice crop growth monitoring,'' \emph{ISPRS Annals of Photogrammetry, Remote Sensing and Spatial Information Sciences}, vol. X-4, pp. 385--395, 2022.

\bibitem{zhang2022comparison}
C.~Zhang and J.~M. Kovacs, ``Comparison of vegetation indices acquired from rgb and multispectral uav data,'' \emph{IEEE Transactions on Geoscience and Remote Sensing}, vol.~60, pp. 1--10, 2022.

\bibitem{huang2021commentary}
S.~Huang, L.~Tang, J.~P. Hupy, Y.~Wang, and G.~Shao, ``A commentary review on the use of normalized difference vegetation index (ndvi) in the era of popular remote sensing,'' \emph{Journal of Forestry Research}, vol.~32, no.~1, pp. 1--6, 2021.

\bibitem{anisa2020uav}
M.~N. Anisa, R.~Hernina \emph{et~al.}, ``Uav application to estimate oil palm trees health using visible atmospherically resistant index (vari)(case study of cikabayan research farm, bogor city),'' in \emph{E3S Web of Conferences}, vol. 211.\hskip 1em plus 0.5em minus 0.4em\relax EDP Sciences, 2020, p. 05001.

\end{thebibliography}

\end{document}